\documentclass[12pt,preprint]{aastex} 
\usepackage{epsfig,lscape}

\slugcomment{KSUPT-04/1 \hspace{0.5truecm} March 2004}

\begin{document}

\title{Constraints on Scalar-Field Dark Energy from the Cosmic Lens All-Sky
Survey Gravitational Lens Statistics}

\author{Kyu-Hyun Chae,\altaffilmark{1} Gang Chen,\altaffilmark{2} 
        Bharat Ratra,\altaffilmark{2} and Dong-Wook Lee\altaffilmark{1}}
\altaffiltext{1}{Department of Astronomy and Space Sciences, Sejong University,
                  98 Gunja-dong, Gwangjin-Gu, Seoul 143-747, Republic of Korea;
\mbox{chae@arcsec.sejong.ac.kr}, \mbox{dwlee@arcsec.sejong.ac.kr}}
\altaffiltext{2}{Department of Physics, Kansas State University, 116 
                 Cardwell Hall, Manhattan, KS 66506;
\mbox{chengang@phys.ksu.edu}, \mbox{ratra@phys.ksu.edu}}

\begin{abstract} 
We use the statistics of strong gravitational lensing based on the Cosmic 
Lens All-Sky Survey (CLASS) data to constrain cosmological parameters in 
a spatially-flat, inverse power-law potential energy density, scalar-field 
dark energy cosmological model. The lensing-based constraints are consistent 
with, but weaker than, those derived from Type~Ia supernova 
redshift-magnitude data, and mildly favor the Einstein cosmological constant
limit of this dark energy model.
 
\end{abstract}

\keywords{cosmology: cosmological parameters---cosmology: 
observation---large-scale structure of the universe---gravitational lensing}

\section{Introduction} 
Recent cosmological measurements strengthen the evidence from Type~Ia
supernova redshift-magnitude measurements (Riess et al.~1998; Perlmutter 
et al.~1999) that the energy density of the current universe is dominated by
Einstein's cosmological constant $\Lambda$, or by a dark energy term in the
cosmic stress-energy tensor that only varies slowly with time and space and so
acts like $\Lambda$. These measurements include: (1) more recent Type~Ia 
supernova redshift-magnitude measurements (see, e.g., Knop et al.~2003; 
Barris et al.~2004); (2) the space-based Wilkinson Microwave Anisotropy Probe 
(WMAP) measurement of the cosmic microwave background (CMB) anisotropy, with 
some input from other measurements (see, e.g., Page et al.~2003; Spergel et 
al.~2003; Scranton et al.\ 2003); and (3) other measurements of CMB anisotropy,
which indicate the universe is close to spatially flat 
(see, e.g., Podariu et al.~2001b; 
Durrer, Novosyadlyj, \& Apunevych 2003; Melchiorri \& \"Odman 2003), in 
combination with the continuing strong evidence for low non-relativistic matter
density (Chen \& Ratra 2003b and references therein). See Peebles \& Ratra 
(2003), Padmanabhan (2003), Bernardeau (2003), Steinhardt (2003), and 
Carroll (2004) for reviews of the current state of affairs.\footnote{
Specific dark energy models and observational measurements are considered 
in Munshi, Porciani, \& Wang (2003), Barreiro 
et al.~(2003), Mainini et al.~(2003), Lima, Cunha, \& Alcaniz (2003), 
Silva \& Bertolami (2003), Amendola et al.~(2003), Linder \& Jenkins (2003), 
Makler, Oliveira, \& Waga (2003), Bean \& Dor\'e (2003), {\L}okas, Bode, 
\& Hoffman (2003), Alam et al.~(2003), Choudhury \& Padmanabhan (2003), 
 Zhu \& Fujimoto (2004), and Macci\'o (2004), from which the earlier 
literature may be accessed.}

While Einstein's $\Lambda$ was the first example of dark energy, nowadays
much attention is focused on scalar field models in which the energy 
density slowly decreases with time and so behaves like a time-variable
$\Lambda$ (see, e.g., Peebles 1984; Peebles \& Ratra 1988, 2003;
Padmanabhan 2003; Steinhardt 2003; Carroll 2004). A simple scalar field 
dark energy model has scalar field $(\phi)$ potential energy density 
$V(\phi) \propto \phi^{-\alpha}$ at low redshift, with $\alpha > 0$ 
(see, e.g., Peebles \& Ratra 1988; Ratra \& Peebles 1988). 
Podariu \& Ratra (2000), Waga \& Frieman (2000), and Gott et al.~(2001) 
examine constraints on this model using Type~Ia supernova 
redshift-magnitude data. They find that a broad range of $\alpha$ 
is consistent with the supernova data.\footnote{
The proposed SNAP space mission (see http://snap.lbl.gov/, and Schubnell 2003 
and Annis et al.~2003) will provide significantly tighter constraints on such 
models (Podariu, Nugent, \& Ratra 2001a; Ericksson \& Amanullah 2002; 
Caresia, Matarrese, \& Moscardini 2003; Wang \& Mukherjee 2003, and 
references therein). Mukherjee et al.~(2003a, 2003b), Spergel et al.~(2003), 
Caldwell \& Doran (2003), Weller \& Lewis (2003), Giovi, Baccigalupi, \& 
Perrotta (2003), and references therein, discuss constraints on scalar field 
and related dark energy models from CMB anisotropy measurements; upcoming 
WMAP and other CMB data will improve these constraints.}  
  
It is important that these dark energy models be tested by other independent 
methods. The redshift--angular-size test is one option. Indications from 
current data, while not as compelling as those discussed above, are 
consistent with a significant dark energy density at low redshift (see, e.g., 
Daly \& Guerra 2002; Zhu \& Fujimoto 2002; Chen \& Ratra 2003a; Podariu 
et al.~2003; Jain, Dev, \& Alcaniz 2003; Jackson 2003). Future 
higher-quality data should turn this into a much more precise cosmological 
test. The redshift-counts test also appears to be on the verge of becoming 
a very promising test (see, e.g., Newman \& Davis 2000; Huterer \& Turner 
2001; Podariu \& Ratra 2001; Levine, Schulz, \& White 2002). Statistical 
analyses of strong gravitational lensing can be used 
to provide constraints on cosmological parameters. Fukugita, Futamase, \& 
Kasai (1990) and Turner (1990) note that the rate of gravitational lensing 
increases rapidly with increasing $\Lambda$. Ratra \& Quillen (1992) and 
Waga \& Frieman (2000) study gravitational lensing in the inverse power-law 
potential scalar field dark energy model. 

The recently completed Cosmic Lens All-Sky Survey (CLASS) is the largest 
uniform survey for strong lensing (Myers et al.\ 2003; Browne et al.\ 2003). 
The survey has discovered 22 cases of multiple-imaging (that are induced by
galaxy-scale lens potentials) out of $\sim 16,500$ extragalactic radio sources.
A subsample of 8958 sources containing 13 multiply-imaged sources 
satisfy well-defined observational selection criteria and is referred
to as the CLASS statistical sample (Browne et al.\ 2003). The CLASS statistical
sample has been used to constrain cosmological parameters (see, e.g., Chae 
et al.\ 2002; Chae 2003; Kuhlen, Keeton, \& Madau 2004; Mitchell et al.\ 2004) 
as well as to constrain global properties of galaxies (Chae 2003; Davis, 
Huterer, \& Krauss 2003) and galaxy evolution (Chae \& Mao 2003). The 
lensing-based constraints on cosmological parameters are consistent with those 
based on Type~Ia supernova magnitude-redshift data but have larger statistical 
errors.

In this work we use the CLASS statistical sample to constrain the 
inverse power-law potential scalar-field dark energy model (Peebles \& Ratra 
1988). In linear perturbation theory, a scalar field is mathematically 
equivalent to a fluid with time-dependent equation of state parameter 
$w = p/\rho$ and speed of sound squared $c_s^2 = \dot p/\dot\rho$, where $p$
and $\rho$ are the pressure and energy density, and the dot denotes a time 
derivative (see, e.g., Ratra 1991). The XCDM parametrization of this dark 
energy model approximates $w$ as a constant, which is accurate during the 
radiation and matter dominated epochs but not in the current, scalar-field 
dark energy dominated epoch. This XCDM approximation thus leads to 
inaccurate predictions for the gravitational lensing considered here, which 
probes the low redshift universe. We emphasize, however, that unlike a lot
of earlier work, we do not work in the XCDM approximation, instead we 
explicitly integrate the scalar-field dark energy equations of motion.

In $\S$2 we summarize the data and method used. Results are presented and
discussed in $\S$3. 

\section{Data and Method}

We use the data listed in Chae (2003) except for the following 
modifications owing to the very recent spectroscopic observations
of several CLASS lens systems by McKean et al.\ (2004). The 13 lens systems
in the CLASS statistical sample (Table~1 of Chae 2003) are 0218+357, 
0445+123, 0631+519, 0712+472, 0850+054, 1152+199, 1359+154, 1422+231, 1608+656,
1933+503, 2045+265, 2114+022, and 2319+051. From McKean et al.\ (2004) we 
adopt the following lens redshifts $z_l =  0.558$, 0.620, and 0.588 
respectively for 0445+123, 0631+519, and 0850+054. We also use the finding by 
McKean et al.\ (2004) that the lenses for 0445+123 and 0631+519 are 
early-type galaxies while that for 0850+054 is a spiral-type galaxy. 

Sheth et al.\ (2003) directly estimate the velocity (dispersion) function (VF)
of early-type galaxies based on the Sloan Digital Sky Survey (SDSS) data; a 
correction to their normalization is reported in Mitchell et al.\ (2004). 
It would be desirable to use the Sheth et al.\ (2003) VF for lensing analyses.
However, we find a posteriori that the maximum likelihood for the Sheth et al.\
(2003) VF is far worse than that for the Chae (2003) inferred VF based on the
Second Southern Sky Redshift Survey (SSRS2). This implies that the SSRS2 VF
is more consistent with the image separations in the CLASS statistical sample
(Chae 2004, in preparation). In this work we use the SSRS2 VF as in Chae et 
al.\ (2002), Chae (2003), and Chae \& Mao (2003). 

We use the method of statistical analysis of lensing described in Chae (2003)
except we now work in the spatially-flat scalar-field dark energy cosmological
model (Peebles \& Ratra 1988). In particular, as in Chae (2003) we assume that
the comoving number density of early-type galaxies is constant from $z \sim 1$
to the present epoch and the characteristic velocity dispersion for 
$0.3 \la z \la 1$ is not assumed known a priori but determined from the
image-splitting sizes of the multiply-imaged systems.\footnote{Chae \& Mao 
(2003) find that if a spatially flat universe with $\Omega_{\rm m,0} = 0.3$ 
and Einstein's $\Lambda$ is assumed, the CLASS data are consistent with
non-evolution of early-type galaxies since $z \sim 1$.}
Here we briefly review essential concepts in lensing statistics (Chae 2003) and
the spatially-flat scalar-field dark energy cosmological model (Peebles \& 
Ratra 1988). Let the differential probability for a cosmologically distant 
source to be multiply-imaged with image-splitting size $\Delta \theta$ to 
$\Delta \theta + d(\Delta \theta)$ by a lens at redshift $z$ to $z + dz$ be
$\delta p$ and the probability of multiple-imaging be the integral of 
$\delta p$ over $\Delta \theta$ and $z$. Then for a statistical sample that
contains $N_{\rm L}$ lensed sources and $N_{\rm U}$ unlensed sources, the
likelihood $L$ of the observation given the statistical lensing
model including the background cosmology is
\begin{equation}
\ln L = \sum_{k=1}^{N_{\rm U}} \ln (1 - p_k) + 
                  \sum_{l=1}^{N_{\rm L}} \ln \delta p_l.
\end{equation}
The differential and integrated lensing probabilities depend both on
the properties of galaxies and on the underlying cosmological model through
proper time element and angular-diameter distances (see, e.g., Chae 2003), 
so that the above likelihood has dependence on cosmological parameters under
consideration. 

The action for the scalar-field model of Peebles \& Ratra (1988) is given by
\begin{equation}
S = \int d^4x \sqrt{-g} \left[ \frac{m_p^2}{16\pi}\left(-R
  + \frac{1}{2} g^{\mu \nu} \partial_{\mu} \phi \partial_{\nu} \phi
     - \frac{1}{2} \kappa m_p^2 \phi^{-\alpha} \right) + 
     \mathcal{L}  \right],
\end{equation}
where `natural units' (i.e.\ $\hbar = c = 1$) are adopted, the Planck mass
$m_p = G^{-1/2}$ ($G$ is Newton's gravitational constant), $\mathcal{L}$ is
the Lagrangian density for matter and radiation, and $\kappa > 0$ and 
$\alpha > 0$ are the parameters characterizing the scalar-field inverse 
power-law potential energy density. For a spatially flat cosmological model
equation~(2) yields the following equations of motion:
\begin{eqnarray}
\ddot{\phi} + 3 \frac{\dot{a}}{a} \dot{\phi} - 
\frac{\kappa \alpha}{2} m_p^2 \phi^{-(\alpha+1)} & = & 0, \nonumber \\
\left(\frac{\dot{a}}{a}\right)^2 & = &\frac{8\pi}{3 m_p^2} (\rho +\rho_\phi),
      \nonumber \\
\rho_\phi & = & \frac{m_p^2}{32\pi}[(\dot{\phi})^2+\kappa m_p^2 
  \phi^{-\alpha}],  \nonumber \\
p_\phi & = & \frac{m_p^2}{32\pi}[(\dot{\phi})^2- \kappa m_p^2 \phi^{-\alpha}],
\end{eqnarray}
where dots denote derivatives with respect to time, $a = a(t)$ is the 
cosmological scale factor, and $\rho_\phi$ and $p_\phi$ are respectively 
the energy density and pressure of $\phi$.
Cosmological quantities in the above model are 
computed through a combination of numerical integration, 
tabulation, and interpolation. Specifically, we numerically integrate the 
equations of motion given by equation~(3) to compute $|d \ell / dz|$
where $\ell$ is the proper time normalized by the Hubble time (see \S 2.1.2 of 
Chae 2003). We compute and tabulate the values of $|d \ell / dz|$ in the
3-dimensional grid spanned by $\Omega_{\rm m,0}$, $\alpha$, and $z$. Then 
the value of $|d \ell / dz|$ for any $\Omega_{\rm m,0}$, $\alpha$, and $z$ is
obtained by interpolation and the angular-diameter distance between two 
redshifts is obtained as usual by the trivial numerical integration of
$(1+z) |d \ell / dz|$. 

\section{Results and Discussion}

Figure 1 shows the CLASS lensing-based constraints on the parameters of the 
spatially-flat inverse power-law potential scalar field dark energy 
cosmological model. The likelihood is maximized for $\Omega_{\rm m,0} = 
0.34$ and $\alpha = 0$, i.e., a conventional cosmological constant. At 68\% 
confidence, $\alpha < 2.7$ and $0.18 < \Omega_{\rm m,0} < 0.62$.\footnote{
If the SDSS measured
VF (Sheth et al.\ 2003; Mitchell et al.\ 2004) were used instead of the
SSRS2 inferred VF (Chae 2003), these ranges would be narrower and the
maximum likelihood estimate of $\Omega_{\rm m,0}$ would be $\sim 0.2$.
See \S 2 and Chae (2004, in preparation) for why we choose to use the
SSRS2 VF.} However, at 95\% confidence 
both $\alpha = 8$ and $\Omega_{\rm m,0}=1$ are allowed. As mentioned in \S 2,
these results are based on the assumption that the comoving number density of
early-type galaxies is unchanged from $z \sim 1$. However, if there were fewer
early-type galaxies at intermediate redshifts compared with the present epoch,
the maximum likelihood estimate of $\Omega_{\rm m,0}$ would become lower and 
the confidence ranges for $\alpha$ and $\Omega_{\rm m,0}$ would become 
narrower. The above results are consistent with, but not as constraining as, 
those derived from Type~Ia supernova redshift-magnitude data (Podariu \& 
Ratra 2000; Waga \& Frieman 2000). They are also consistent with, but more
constraining than, those determined using measurements of angular size as
a function of redshift (Chen \& Ratra 2003a; Podariu et al.~2003).

It is interesting to note that various disparate data sets give consistent
constraints on the inverse power-law potential energy density scalar-field 
dark energy model that weakly favor the conventional cosmological constant over
a dynamical scalar field dark energy. However, current results are tentative 
and future much larger data sets are required to resolve this issue. 
Future lensing data (e.g.,\ CLASS2; see \S 6 of Chae 2003) would be valuable 
in this respect and are eagerly anticipated.

\bigskip

We thank the anonymous referee for constructive comments.
KHC and DWL acknowledge support from the ARCSEC of KOSEF.
GC and BR acknowledge support from NSF CAREER grant AST-9875031
and DOE EPSCoR grant DE-FG02-00ER45824.


\begin{figure}
\centerline{\epsfig{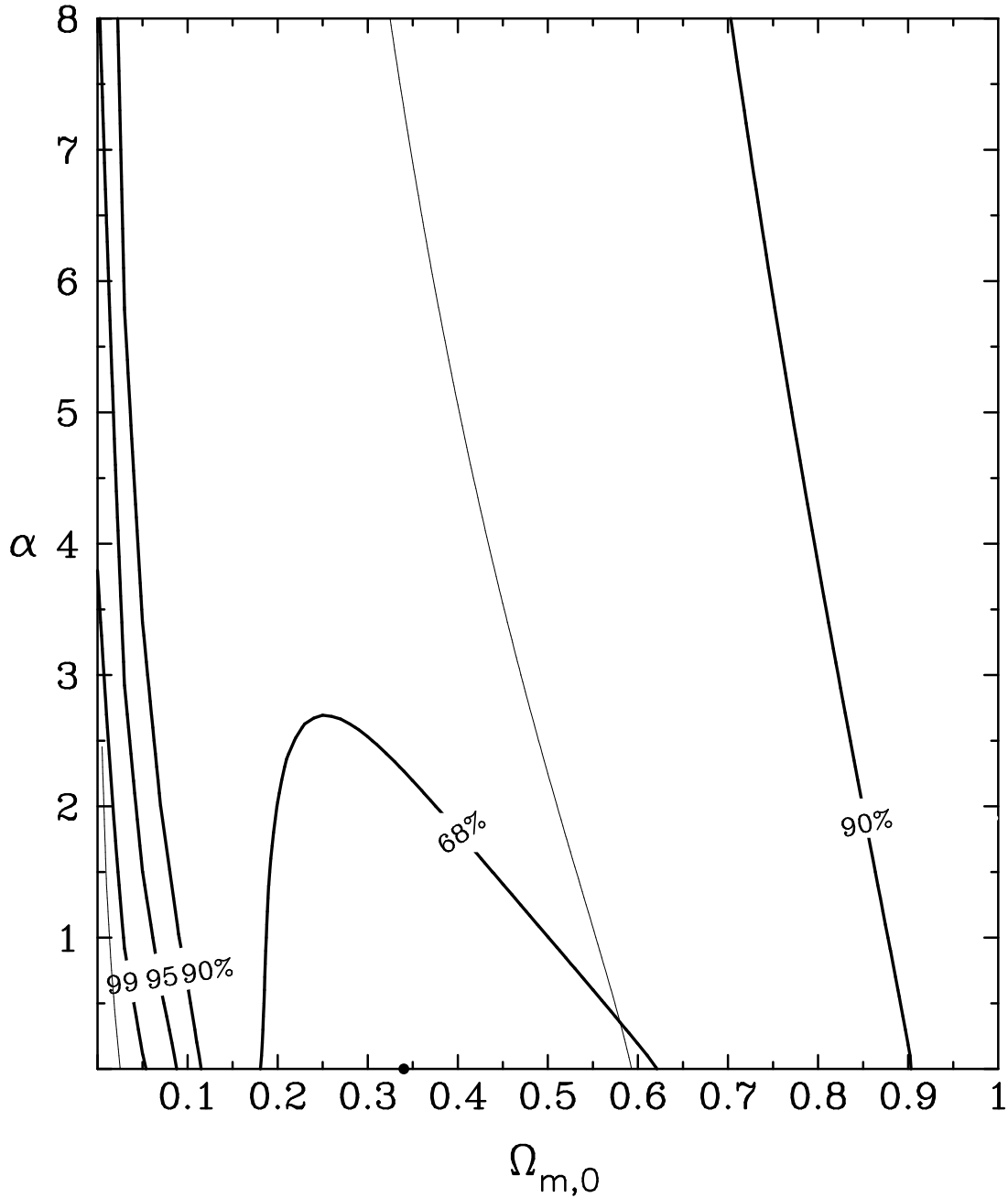}}
\caption{Contours of 68, 90, 95, and 99\% confidence based on a
likelihood ratio test (only the left parts 
of the last two are shown, near the left hand edge of the plot) for the 
spatially-flat scalar-field dark energy model with potential energy density 
$V(\phi) \propto \phi^{-\alpha}$ at low redshift. Black dot on the horizontal
axis near non-relativistic matter density parameter $\Omega_{\rm m,0} = 0.34$ 
denotes where the likelihood is maximized. Overplotted thin lines represent 
68\% confidence limit by recent redshift--angular-size data (Chen \& Ratra 
2003a), which are given for comparison. Here the confidence contours are based
on the same likelihood ratio test as for the lensing data. These contours are, 
however, different from those of Chen \& Ratra (2003a) because they are from
fractions of the integrated likelihood over the whole plane assuming the prior
that the likelihood is zero outside the range  $0 < \Omega_{\rm m,0} <1$ 
and $0 < \alpha < 8$.}
\label{f1}
\end{figure} 

\end{document}